\documentclass[12pt]{article}
\usepackage[centertags]{amsmath}
\usepackage{amssymb,amsfonts}
\usepackage[square,comma]{natbib}
\usepackage{latexsym}
\newcommand{\be}{\begin{equation}} \newcommand{\ee}{\end{equation}}
\newcommand{\bea}{\begin{eqnarray}} \newcommand{\eea}{\end{eqnarray}}
\newcommand{\bse}{\begin{subequations}} \newcommand{\ese}{\end{subequations}}
\newcommand{\n}{\nonumber}
\begin{document}
\title{\textbf{Generalised compact spheres in electric fields}}
\author{S. D. Maharaj\thanks{eMail:
\texttt{maharaj@ukzn.ac.za}}\; and {K. Komathiraj\thanks{Permanent
address: Department of Mathematical Sciences, South Eastern
University, Sammanthurai, Sri Lanka.}}\\
Astrophysics and Cosmology Research Unit,\\
School of Mathematical Sciences, University of KwaZulu-Natal,\\
Private Bag X54001, Durban 4000, South Africa.}
\date{}
\maketitle

\begin{abstract}
We present exact solutions to the Einstein-Maxwell system of
equations in spherically symmetric gravitational fields with a
specified form of the electric field intensity. The condition of
pressure isotropy yields a difference equation with variable,
rational coefficients. In an earlier treatment this condition was
integrated by first transforming it to a hypergeometric equation. We
demonstrate that it is possible to obtain a more general class of
solutions to the Einstein-Maxwell system both in the form of special
functions and elementary functions. Our results contain particular
solutions found previously including models of charged relativistic
spheres and uncharged neutron star models.
\end{abstract}

\section{Introduction}
Exact solutions of the Einstein-Maxwell system are important in the
description of relativistic astrophysical processes. In the presence
of charge the gravitational collapse of a spherically symmetric
matter configuration to a point singularity may be avoided
\cite{Kra}. Einstein-Maxwell solutions are important in studies
involving the cosmic censorship hypothesis and the formation of
naked singularities \cite{Jos}. The presence of the electromagnetic
field affects the values of the surface redshifts, luminosities and
the maximum mass for charged stars as demonstrated by Sharma
\emph{et al} \cite{ShMuMa}, Ivanov \cite{Iva} and others. The
analysis of Patel and Koppar \cite{PaKo}, Patel \emph{et al}
\cite{PaTiSa}, Tikekar and Singh \cite{TiSi}, Gupta and Kumar
\cite{GuKu}, and Mukherjee \cite{Muk} have shown that it is feasible
to model charged superdense neutron stars with densities in the
range of 10$^{14}$ g cm$^{-3}$; these models generate bounds for the
surface readshift, luminosity and stellar mass which are consistent
with observation. Sharma \emph{et al} \cite{ShKaMu} studied cold
compact spheres, Tikekar and Jotania \cite{TiJo} analysed strange
matter, Sharma and Mukherjee \cite{ShMu} considered the equation of
state of a compact X-ray binary pulsar Her X-1, and Sharma and
Mukherjee \cite{Sha} investigated stars composed of quark-diquark
particles which are consistent with the presence of an
electromagnetic field. Einstein-Maxwell solutions may be applied to
the core envelope models of Thomas \emph{et al} \cite{ThRaVi},
Tikekar and Thomas \cite{TiTh}, and Paul and Tikekar \cite{PaTi}
with an inner core surrounded by an outer layer. These references
provide a sample as to why the Einstein-Maxwell system, describing
the interior of a charged star, has attracted the attention of many
researchers.

Our intention in this paper is two-fold. Firstly, we seek to model a
charged relativistic sphere which is physically acceptable. We
require that the gravitational, electromagnetic and matter variables
are finite, continuous and well behaved in the stellar interior, the
interior metric should match smoothly with the exterior
Reissner-Nordstrom metric, the speed of sound is less that the speed
of light, and the solution is stable with respect to radial
perturbations. Secondly, we seek to regain an uncharged solution of
Einstein equations which satisfy the relevant physical criteria when
the electric field vanishes. This ensures that a neutral
relativistic star is regainable as a stable equilibrium state. We
seek to construct a model, with limiting uncharged stars as known
exact solutions, that exactly satisfies the Einstein-Maxwell
systems. This ideal is not easy to achieve in practice and only a
few examples with the required two features have been found thus
far. Recently Hansraj and Maharaj \cite{HaMa}, Thirukkanesh and
Maharaj \cite{ThMa}, and Komathiraj and Maharaj \cite{KoMa,KoMah}
found charged relativistic fluid spheres which regain neutral
compact stars for particular parameter values.

Our objective is to find new solutions of the Einstein-Maxwell
system that satisfy the physical criteria given above, and
necessarily contains a neutral stellar solution. Our approach here
complements  the approach of Thirukkanesh and Maharaj \cite{ThMa}
who presented exact solutions to the Einstein-Maxwell equations;
this family of solutions can be written in terms of elementary
functions and reduces to the well known uncharged stellar model of
Durgapal and Bannerji \cite{DuBa}. The approach in \cite{ThMa}
introduces a transformation that reduces the condition of pressure
isotropy to a hypergeometric equation. The transformation utilised
in \cite{ThMa} does produce new exact models but restricts the
classes of solutions that are possible because of constraints placed
on particular parameters. In this paper we do not transform the
condition of pressure isotropy to the hypergeometric equation but
are still in a position to integrate the field equations. A new
class of Einstein-Maxwell solutions are found that contain familiar
uncharged models which are regainable for different choices of the
metric function and the electric field. In Section 2, we express the
Einstein-Maxwell field equations for the static spherically
symmetric line element as an equivalent set of differential
equations utilising a transformation from \cite{DuBa}. We choose
particular forms for one of the gravitational potentials and the
electric field intensity. This enables us to obtain the condition of
pressure isotropy in the remaining gravitational potential which
facilitates the integration process. In Section 3, we assume a
solution in a series form which yields recurrence relations, which
we manage to solve from first principles. It is then possible to
exhibit exact solutions to the Einstein-Maxwell system. Solutions in
terms of elementary functions are possible for particular parameter
values. In Section 4, we present two linearly independent classes of
solutions as combination of polynomials and algebraic functions. In
addition we show that it is possible to express the general solution
of the Einstein-Maxwell system in terms of elementary functions. We
regain some solutions found previously from our general solutions in
Section 5. In Section 6, we summarise the results found and present
plots illustrating the behaviour of the matter variables.

\section{The isotropic equations}
We take the line element for static spherically symmetric
spacetimes to be  \be \label{eq:1}
ds^{2}=-e^{2\nu(r)}dt^{2}+e^{2\lambda(r)}dr^{2}+r^{2}(d\theta^{2}+\sin^{2}\theta
d\phi^{2})\ee where $\nu(r)$ and $\lambda(r)$ are arbitrary
functions. For charged perfect fluids, the Einstein-Maxwell field
equations can be expressed as follows  \bse\label{eq:2}\bea
\label{eq:2a}\frac{1}{r^{2}}[r(1-e^{-2\lambda})]^{\prime}&=&\rho+\frac{1}{2}E^{2}\\\n\\
\frac{-1}{r^{2}}(1-e^{-2\lambda})+\frac{2\nu^\prime}{r}e^{-2\lambda}&=&p-\frac{1}{2}E^{2}\\\n\\
e^{-2\lambda}\left(\nu^{\prime\prime}+{\nu^\prime}^2+\frac{\nu^\prime}{r}-\nu^\prime\lambda^\prime-\frac{\lambda^\prime}{r}\right)&=&p+\frac{1}{2}E^{2}\\\n\\
\label{eq:2d}\sigma&=&\frac{1}{r^{2}}e^{-\lambda}(r^{2}E)^\prime\eea\ese
for the geometry described by (\ref{eq:1}). The energy density
$\rho$ and the pressure $p$ are measured relative to the comoving
fluid 4-velocity $u^{a}=e^{-\nu}\delta^{a}_{0}$, $E$ is the
electric field intensity, $\sigma$ is the proper charge density,
and primes denote differentiation with respect to $r$. In the
system (\ref{eq:2a})-(\ref{eq:2d}) we are using units where the
coupling constant $\frac{8\pi G}{c^{4}}=1$ and the speed of light
$c=1$. The field equations (\ref{eq:2a})-(\ref{eq:2d}) govern the
behaviour of the gravitational field for a charged perfect fluid.
When $E=0$ we regain Einstein equations for a neutral fluid.

An equivalent form of the field equations is generated if we
introduce the transformation  \be \label{eq:3}
A^{2}y^{2}(x)=e^{2\nu(r)},~~~Z(x)=e^{-2\lambda(r)},~~~x=Cr^{2}\ee
where $A$ and $C$ are arbitrary constants. Under the
transformation (\ref{eq:3}), the system (\ref{eq:2}) becomes
 \bse\label{eq:4} \bea
\label{eq:4a}\frac{1-Z}{x}-2\dot{Z}&=&\frac{\rho}{C}+\frac{E^{2}}{2C}\\\n\\
\label{eq:4b}4Z\frac{\dot{y}}{y}+\frac{Z-1}{x}&=&\frac{p}{C}-\frac{E^{2}}{2C}\\\n\\
\label{eq:4c}4Zx^{2}\ddot{y}+2\dot{Z}x^{2}\dot{y}+\left(\dot{Z}x-Z+1-\frac{E^{2}x}{C}\right)y&=&0\\\n\\
\label{eq:4d}\frac{\sigma^{2}}{C}&=&\frac{4Z}{x}(x\dot{E}+E)^{2}\eea\ese
where dots denote differentiation with respect to the variable
$x$. The particular representation of the Einstein-Maxwell system
as given in (\ref{eq:4a})-(\ref{eq:4d}) may be easier to integrate
in certain situation as demonstrated by Hansraj and Maharaj
\cite{HaMa} and Komathiraj and Maharaj \cite{KoMah} among others.
To integrate the system (\ref{eq:4}) it is necessary to specify
two of the variables $y,~Z,~\rho,~p$ or $E$. In our approach we
choose $Z$ and $E$ on physical grounds. The remaining quantities
are then obtained from the rest of the system (\ref{eq:4}).

In the integration procedure we make the choice
\be\label{eq:5}Z=\frac{1+k x}{1+x}\ee where $k$ is a real constant.
For the choice (\ref{eq:5}) the line element (\ref{eq:1}) becomes
\be
ds^{2}=-A^{2}y^{2}dt^{2}+\frac{1+x}{4Cx(1+kx)}dx^{2}+\frac{x}{C}(d\theta^{2}+\sin^{2}\theta
d\phi^{2})\n\ee and we need to find the function $y(x)$. The form
(\ref{eq:5}) ensures that the metric function behaves as
\[ e^{2\lambda} = 1 + O(r^2) \]
near $r=0$. In fact this is a sufficient condition for a static
fluid sphere to be regular at the centre as pointed out by Maartens
and Maharaj \cite{MaMa}. The choice (\ref{eq:5}) ensures that the
function is finite at the centre, remains regular and is well
behaved in the stellar interior. This form contains particular
solutions studied previously; for example when $k= -\frac12$ we
regain the neutron star model of Durgapal and Bannerji \cite{DuBa}.
The choice (\ref{eq:5}) ensures that charged spheres found as exact
solutions of the Einstein-Maxwell system will contain physically
reasonable uncharged models when $E=0$.

On substituting (\ref{eq:5}) in (\ref{eq:4c}) we obtain
\be\label{eq:6}4(1+kx)(1+x)\ddot{y}+2(k-1)\dot{y}+\left[(1-k)-\frac{E^{2}(1
+x)^{2}}{Cx}\right]y=0\ee which is the condition of pressure
isotropy. The solution of the Einstein-Maxwell system (\ref{eq:4}),
for the form (\ref{eq:5}), depends on the integrability of
(\ref{eq:6}). It is necessary to specify the electric field
intensity $E$ to integrate (\ref{eq:6}). A variety of choices for
$E$ is possible but only a few are physically reasonable which
generate closed form solutions. We can reduce (\ref{eq:6}) to
simpler form if we let
 \be\label{eq:7}\frac{E^{2}}{C}=\frac{\alpha
x}{(1+x)^{2}} \ee where $\alpha$ is a constant. The form for $E^{2}$
in (\ref{eq:7}) vanishes at the centre of the star, and remains
continuous and bounded in the interior of the star for a wide range
of values of the parameter $\alpha$. Thus the choice of $E^2$ is
physically reasonable and useful in the gravitational analysis of
charged spheres.
 When $\alpha=0$ there is no
charge, and we regain uncharged stellar solutions such as the
neutron star model of Durgapal and Bannerji \cite{DuBa} for the
parameter value $k= -\frac12$. Note that the same form for the
electric field intensity (\ref{eq:7}) was utilised by Hansraj and
Maharaj \cite{HaMa} and Thirukkanesh and Maharaj \cite{ThMa}. For
this form of $E^2$ a comprehensive physical analysis was performed
by Hansraj and Maharaj \cite{HaMa} for charged spheres when $k=0$.
They demonstrated that all criteria for physical acceptability were
satisfied including the fact that the speed of sound is less that
the speed of light. We expect that this form for the electric field
intensity, with the generalised potential (\ref{eq:5}), will provide
a more general class of charged relativistic spheres with desirable
physical features.

Upon substituting the choice (\ref{eq:7}) in equation (\ref{eq:6})
we obtain
\be\label{eq:8}4(1+kx)(1+x)\ddot{y}+2(k-1)\dot{y}+(1-k-\alpha)y=0\ee
which is the master equation for the system (\ref{eq:4}).
Thirukkanesh and Maharaj \cite{ThMa} introduced the transformation
\be 1+x=KX,~~~~~K=\frac{k-1}{k},~~~~~Y(X)=y(x)\n\ee so that
(\ref{eq:8}) becomes \be
4X(1-X)\frac{d^{2}Y}{dX^{2}}-2\frac{dY}{dX}+(K+\tilde{\alpha})Y=0,~~~\tilde{\alpha}=\frac{\alpha}{k}\n\ee
which is a special case of the hypergeometric equation. It is
possible to integrate this hypergeometric equation. However it is
important to note that the possible solutions are restricted as
$k\neq0$ and $k\neq1$ because of the transformation used. We
demonstrate in Section 3 that we can accommodate $k=0$ and $k=1$ in
a wider class of solutions.
\section{Master Equation}
It is convenient to introduce the new variable $z=1+x$ in
(\ref{eq:8}) to yield \be\label{eq:9}
4z(1-k+kz)\frac{d^{2}\tilde{y}}{dz^{2}}+2(k-1)\frac{d\tilde{y}}{dz}+(1-k-\alpha)\tilde{y}=0\ee
where we have set $\tilde{y}=y(z)$. Two categories of solution are
possible for $k=1$ and $k\neq 1$. \\\\
\textbf{Case I} : $k=1$\\
In this case (\ref{eq:9}) becomes the Euler-Cauchy equation with
solution \be \label{eq:9a}
y=c_{1}(1+x)^{(1+\sqrt{1+\alpha})/2}+c_{2}(1+x)^{(1-\sqrt{1+\alpha})/2}\ee
where $c_{1},~ c_{2}$ are constants. From
(\ref{eq:4a})-(\ref{eq:4d}), (\ref{eq:5}) and (\ref{eq:9a}) we can
show that the solution to the Einstein-Maxwell system, for our
specific choices of $Z$ and $E$,  becomes \bse\bea
\label{eq:9a1}e^{2\lambda}&=&1\\
e^{2\nu}&=&A^{2}[c_{1}(1+x)^{(1+\sqrt{1+\alpha})/2}+c_{2}(1+x)^{(1-\sqrt{1+\alpha})/2}]^{2}\\
\rho&=&-\frac{\alpha Cx}{2(1+x)^{2}}\\
p&=&\frac{2C[c_{1}(1+\sqrt{1+\alpha})(1+x)^{\sqrt{1+\alpha}}+c_{2}(1-\sqrt{1+\alpha})]}{(1+x)[c_{1}(1+x)^{\sqrt{1+\alpha}}+c_{2}]}+\frac{\alpha Cx}{2(1+x)^{2}}\\
\label{eq:9a5}E^{2}&=&\frac{\alpha Cx}{(1+x)^{2}}\eea\ese in terms
of the variable $x$. Observe that this case is not regainable from
Thirukkanesh and Maharaj \cite{ThMa} as $k\neq 1$ in their
transformation. We do not pursue this case further as $\rho<
0$.\\\\
\textbf{Case II} : $k\neq1$ \\As the point $z=0$ is a regular
singular point of (\ref{eq:9}), there exist two linearly
independent solutions of the form of a power series with centre
$z=0$. These solutions can be generated using the method of
Frobenius. Therefore we can write \be \label{eq:10}
\tilde{y}=\sum_{i=0}^{\infty}a_{i}z^{i+b},~~~a_{0}\neq 0\ee where
$a_{i}$ are the coefficients of the series and $b$ is a constant.
For a legitimate solution we need to determine the coefficients
$a_{i}$ as well as the parameter $b$. On substituting
(\ref{eq:10}) in to (\ref{eq:9}) we obtain \bea\label{eq:11}
&&2a_{0}b(1-k)(2b-3)z^{b-1}\n\\
&+&\sum_{i=0}^{\infty}\{2(1-k)(i+b+1)(2i+2b-1)a_{i+1}\n\\
&+&[4k(i+b)(i+b-1)+(1-k-\alpha)]a_{i}\}z^{b+i}=0\eea in increasing
powers of $z$. For equation (\ref{eq:11}) to hold for all powers
of $z$ in the interval of convergence we require
\bse\label{eq:12}\bea \label{eq:12a}2a_{0}b(1-k)(2b-3)&=&0\\\n\\
\label{eq:12b}a_{i+1}&=&\frac{4k(i+b)(i+b-1)+(1-k-\alpha)}{2(k-1)(i+b+1)(2i+2b-1)}a_{i},~~~i\geq
0\eea\ese Since $a_{0}\neq0$ and $k\neq1$, we have from
(\ref{eq:12a}) that $b=0$ or $b=\frac{3}{2}$. Equation
(\ref{eq:12b}) is the basic difference equation governing the
structure of the solution. It is possible to express the
coefficient $a_{i}$ in terms of the leading coefficient $a_{0}$ by
establishing a general structure for the coefficients by
considering the leading terms. These coefficients generate the
pattern \be\label{eq:13}
a_{i+1}=\prod_{p=0}^{i}\frac{4k(p+b)(p+b-1)+(1-k-\alpha)}{2(k-1)(p+b+1)(2p+2b-1)}a_{0}\ee
where we have utilised the conventional symbol $\prod$ to denote
multiplication. It is easy to establish that the result
(\ref{eq:13}) holds for all positive integers $p$ using the
principle of mathematical induction.

Now it is possible to generate two linearly independent solutions
to (\ref{eq:9}) with the help of (\ref{eq:10}) and (\ref{eq:13}).
For the parameter value $b=0$ we obtain the first solution \be
\label{eq:14}
\tilde{y}_{1}=a_{0}\left[1+\sum_{i=0}^{\infty}\prod_{p=0}^{i}\frac{4kp(p-1)+(1-k-\alpha)}{2(k-1)(p+1)(2p-1)}z^{i+1}\right]\ee
For the parameter value $b=\frac{3}{2}$ we obtain the second
solution \be \label{eq:15}
\tilde{y}_{2}=a_{0}z^{3/2}\left[1+\sum_{i=0}^{\infty}\prod_{p=0}^{i}\frac{k(2p
+3)(2p+1)+(1-k-\alpha)}{(k-1)(2p+5)(2p+2)}z^{i+1}\right]\ee
Thus we can write the general solution to the differential
equation (\ref{eq:9}), for the choice of the electric field given
in (\ref{eq:7}), as \be\label{eq:16}
\tilde{y}=\tilde{c}_{1}\tilde{y}_{1}+\tilde{c}_{2}\tilde{y}_{2}\ee
where $\tilde{c}_{1},~\tilde{c}_{2}$ are arbitrary constants, and
$\tilde{y}_{1}$ and $\tilde{y}_{2}$ are given in (\ref{eq:14}) and
(\ref{eq:15}) respectively. In terms of the variable $x$ we can
write (\ref{eq:16}) as \bea \label{eq:17}
y&=&c_{1}\left[1+\sum_{i=0}^{\infty}\prod_{p=0}^{i}\frac{4kp(p-1)
+(1-k-\alpha)}{2(k-1)(p+1)(2p-1)}(1+x)^{i+1}\right]\n\\
&+&c_{2}(1+x)^{3/2}\left[1+\sum_{i=0}^{\infty}\prod_{p=0}^{i}\frac{k(2p
+3)(2p+1)+(1-k-\alpha)}{(k-1)(2p+5)(2p+2)}(1+x)^{i+1}\right]\n\\
&=&c_{1}y_{1}(x)+c_{2}y_{2}(x)\eea where we have set
$c_{1}=\tilde{c}_{1}a_{0}$ and $c_{2}=\tilde{c}_{2}a_{0}$ for
simplicity. The  solution to the Einstein-Maxwell system
(\ref{eq:4a})-(\ref{eq:4d}), for our specific choices of $Z$ and
$E$, can now be written as
\bse \label{20} \bea\label{eq:17a}
e^{2\lambda}&=&\frac{1+x}{1+kx}\\
e^{2\nu}&=&A^{2}y^{2}\\
\label{eq:17c}\frac{\rho}{C}&=&\frac{(1-k)(3+x)}{(1+x)^{2}}-\frac{\alpha
x}{2(1+x)^{2}}\\
\frac{p}{C}&=&4\frac{(1+kx)}{(1+x)}\frac{\dot{y}}{y}+\frac{(k-1)}{(1+x)}+\frac{\alpha x}{2(1+x)^{2}}\\
\label{eq:17d}\frac{E^{2}}{C}&=&\frac{\alpha x}{(1+x)^{2}}\eea\ese
in terms of the variable $x$. The form of the exact solution
(\ref{eq:17a})-(\ref{eq:17d}) has a similar structure to the
general solution of Thirukkanesh and Maharaj \cite{ThMa}; however
it is important to realise that our solution is a new result
because the series in (\ref{eq:17}) is different. In addition note
that $k=0$ is allowed in (\ref{eq:17a})-(\ref{eq:17d}) unlike the
result of Thirukkanesh and Maharaj \cite{ThMa}; our result can be
interpreted as a generalisation.
\section{Elementary solutions}
The general solution (\ref{eq:17}) is given in the form of a
series which define special functions. For particular values of
the parameters involved it is possible for the general solution to
be written in terms of elementary functions which is a more
desirable form for the physical description of a charged
relativistic star. We find two linearly independent solutions, in
terms of elementary functions, in this section for the
differential equation (\ref{eq:9}).
\subsection{First solution}
On substituting $b=0$ and setting $1-k-\alpha=-4kn(n-1)$ in
(\ref{eq:12b}) we obtain \be \label{eq:18}
a_{i+1}=\frac{4k}{1-k}\frac{(n-i)(n+i-1)}{(2i+2)(2i-1)}a_{i},~~~i\geq
0 \ee It is easy to see from (\ref{eq:18}) that $a_{n+1}=0$.
Clearly the subsequent coefficients $a_{n+2},~a_{n+3},\\
~a_{n+4},.~.~.$ vanish. Equation (\ref{eq:18}) has the solution
\be \label{eq:19}
a_{i}=-\left(\frac{4k}{1-k}\right)^{i}\frac{n(n-1)(2i-1)(n+i-2)!}{(2i)!(n-i)!}a_{0},~~~0\leq
i \leq n\ee Then from (\ref{eq:10}) (when $b=0$) and
(\ref{eq:19}), we obtain \be \label{eq:20}
\tilde{y}_{1}=a_{0}\left[1-\sum_{i=1}^{n}\left(\frac{4k}{1-k}\right)^{i}\frac{n(n-1)(2i-1)(n+i-2)!}{(2i)!(n-i)!}z^{i}\right]\ee
for $1-k-\alpha=-4kn(n-1)$.

On substituting $b=\frac{3}{2}$ and setting
$1-k-\alpha=-k(2n+3)(2n+1)$ in (\ref{eq:12b}) we obtain
\be\label{eq:21}
a_{i+1}=\frac{4k}{1-k}\frac{(n-i)(n+i+2)}{(2i+5)(2i+2)}a_{i},~~~i\geq
0 \ee We observe from (\ref{eq:21}) that $a_{n+1}=0$. Consequently
the remaining coefficients $a_{n+2},~a_{n+3},\\~a_{n+4},~.~.~.$
vanish. Equation (\ref{eq:21}) may be solved to yield \be
\label{eq:22}
a_{i}=3\left(\frac{4k}{1-k}\right)^{i}\frac{(2i+2)(n+i+1)!}{(n+1)(2i+3)!(n-i)!}a_{0},~~~0\leq
i \leq n \ee Now from (\ref{eq:10}) (when $b=\frac{3}{2}$) and
(\ref{eq:22}), we obtain \be \label{eq:23}
\tilde{y}_{1}=a_{0}z^{3/2}\left[1+3\sum_{i=1}^{n}\left(\frac{4k}{1-k}\right)^{i}\frac{(2i+2)(n+i+1)!}{(n+1)(2i+3)!(n-i)!}z^{i}\right]\ee
for $1-k-\alpha=-k(2n+3)(2n+1)$.

The polynomial and algebraic functions (\ref{eq:20}) and
(\ref{eq:23}) comprise the first solution to the differential
equation (\ref{eq:9}) for appropriate values of $k$ and $\alpha$.

\subsection{Second solution}
We can find the second solution of (\ref{eq:9}) using the method of
reduction of order in principle. However this proves to be difficult
in practice because of the complicated form of the first solution
given in  (\ref{eq:20}) and (\ref{eq:23}). We utilise a
transformation to first simplify (\ref{eq:9}) before seeking the
second solution. We take the second solution of (\ref{eq:9}) to be
of the form \be \label{eq:24} \tilde{y}=(1-k+kz)^{1/2}u(z)\ee when
$u(z)$ is an arbitrary polynomial. Special cases of (\ref{eq:24})
are known to solve (\ref{eq:9}) which motivates the algebraic form
for $\tilde{y}$ as a generic second solution to the differential
equation (\ref{eq:9}). On substituting $\tilde{y}$ in (\ref{eq:9})
we obtain\be\label{eq:25}
4z(1-k+kz)\frac{d^{2}u}{dz^{2}}+[4kz+2(k-1)]\frac{du}{dz}+(1-2k-\alpha)u=0\ee
which is linear differential equation for $u(z)$.

As in Section 4.1 it is possible to find solutions in terms of
polynomials, and product of polynomials with algebraic functions
for $u(z)$ for certain values of the parameters $k$ and $\alpha$.
As the point $z=0$ is a regular singular point of (\ref{eq:25}),
there exist two linearly independent solutions of the form of the
power series with centre $z=0$. Thus we assume \be \label{eq:26}
u=\sum_{i=0}^{\infty}c_{i}z^{i+d}\ee where the constants $c_{i}$
are the coefficients of the series and $d$ is the constant.
Substituting  (\ref{eq:26}) in (\ref{eq:25}) we obtain \bea \label{eq:27} &&2c_{0}d(1-k)(2d-3)z^{d-1}\n\\
&+&\sum_{i=0}^{\infty}\{2(1-k)(i+d+1)(2i+2d-1)c_{i+1}\n\\
&+&[4k(i+d)^{2}+(1-2k-\alpha)]c_{i}\}z^{i+d}=0\eea For equation
(\ref{eq:27}) to hold true for all $z$ we require that
\bse\label{eq:28}\bea
\label{eq:28a}2c_{0}d(1-k)(2d-3)&=&0 \\
\label{eq:28b}c_{i+1}&=&
\frac{4k(i+d)^{2}+(1-2k-\alpha)}{2(k-1)(i+d+1)(2i+2d-1)}c_{i},~~~i
\geq 0\eea\ese Since $c_{0}\neq0$ and $k\neq1$ we have from
(\ref{eq:28a}) that $d=0$ or $d=\frac{3}{2}$. Equation
(\ref{eq:28b}) is the linear recurrence relation governing the
structure of the solution.

On substituting $d=0$ and setting $1-k-\alpha=-k(2n+1)(2n+3)$ in
(\ref{eq:28b}) we obtain \be\label{eq:29}
c_{i+1}=\left(\frac{4k}{1-k}\right)\frac{(n-i+1)(n+i+1)}{(2i+2)(2i-1)}c_{i},~~~i\geq
0\ee From (\ref{eq:29}) we have that $c_{n+2}=0$ and subsequent
coefficients $c_{n+3},~c_{n+4},~c_{n+5},~.~.~.$ vanish. Then
(\ref{eq:29}) has the solution \be\label{eq:30}
c_{i}=-\left(\frac{4k}{1-k}\right)^{i}\frac{(n+1)(2i-1)(n+i)!}{(2i)!(n-i+1)!}c_{0},~~~0\leq
i \leq n+1 \ee Then from (\ref{eq:26}) (when $d=0$) and
(\ref{eq:30}) we obtain \be\label{eq:31}
\tilde{y}_{2}=c_{0}(1-k+kz)^{1/2}\left[1-\sum_{i=1}^{n+1}\left(\frac{4k}{1-k}\right)^{i}\frac{(n+1)(2i-1)(n+i)!}{(2i)!(n-i+1)!}z^{i}\right]\ee
for $1-k-\alpha=-k(2n+1)(2n+3)$.

On substituting $d=\frac{3}{2}$ and setting $1-k-\alpha=-4kn(n-1)$
in (\ref{eq:28b}) we obtain \be\label{eq:32}
c_{i+1}=\left(\frac{4k}{1-k}\right)\frac{(n+i+1)(n-i-2)}{(2i+2)(2i+5)}c_{i},~~~i\geq
0 \ee From  (\ref{eq:32}) we have that $c_{n-1}=0$ and subsequent
coefficients $c_{n},~c_{n+1},~c_{n+2},~.~.~.$ vanish. The equation
(\ref{eq:32}) has the solution \be\label{eq:33}
c_{i}=3\left(\frac{4k}{1-k}\right)^{i}\frac{(2i+2)(n+i)!}{n(n-1)(2i+3)!(n-i-2)!}c_{0},~~~0\leq
i \leq n-2 \ee From (\ref{eq:26}) (when $d=\frac{3}{2}$) and
(\ref{eq:33}) we obtain \be\label{eq:34}
\tilde{y}_{2}=c_{0}(1-k+kz)^{1/2}z^{3/2}\left[1+3\sum_{i=1}^{n-2}\left(\frac{4k}{1-k}\right)^{i}\frac{(2i+2)(n+i)!}{n(n-1)(2i+3)!(n-i-2)!}z^{i}\right]\ee
for $1-k-\alpha=-4kn(n-1)$.

The algebraic solutions (\ref{eq:31}) and (\ref{eq:34}) comprise
the second solution of the differential equation (\ref{eq:9})
which are clearly independent from (\ref{eq:23}) and
(\ref{eq:20}).

\subsection{Elementary functions}
We have obtained one class of polynomial solution (\ref{eq:20})
and three classes of  solutions (\ref{eq:23}), (\ref{eq:31}), and
(\ref{eq:34}) in terms of product of polynomials and algebraic
functions. The polynomial solution (\ref{eq:20}) and the product
of polynomials  with an algebraic function (\ref{eq:23}) generate
the first solution. The second linearly independent solution is
given by  (\ref{eq:31}) and (\ref{eq:34}) which are products of
polynomials and algebraic functions. By collecting these results
we can express the general solution to (\ref{eq:9}) in two
categories. We express the general solution in terms of the
independent variable $x$. From (\ref{eq:20}) and (\ref{eq:34}) we
write the first category as \bea\label{eq:35}
y&=&A\left[1-\sum_{i=1}^{n}\left(\frac{4k}{1-k}\right)^{i}\frac{n(n-1)(2i-1)(n+i-2)!}{(2i)!(n-i)!}(1+x)^{i}\right]\n\\
&+&B(1+kx)^{1/2}(1+x)^{3/2}\left[1+3\sum_{i=1}^{n-2}\left(\frac{4k}{1-k}\right)^{i}\frac{(2i+2)(n+i)!}{n(n-1)(2i+3)!(n-i-2)!}(1+x)^{i}\right]\n\\\eea
for $1-k-\alpha=-4kn(n-1)$. From (\ref{eq:23}) and (\ref{eq:31}),
the second category of solution has the form \bea\label{eq:36}
y&=&A(1+x)^{3/2}\left[1+3\sum_{i=1}^{n}\left(\frac{4k}{1-k}\right)^{i}\frac{(2i+2)(n+i+1)!}{(n+1)(2i+3)!(n-i)!}(1+x)^{i}\right]\n\\
&+&B(1+kx)^{1/2}\left[1-\sum_{i=1}^{n+1}\left(\frac{4k}{1-k}\right)^{i}\frac{(n+1)(2i-1)(n+i)!}{(2i)!(n-i+1)!}(1+x)^{i}\right]\n\\\eea
for $1-k-\alpha=-k(2n+1)(2n+3)$. In the above $A$ and $B$ are
arbitrary constants. Consequently we have demonstrated that
elementary functions can be extracted from the general series
(\ref{eq:17}) by restricting the parameter values $\alpha$ and $k$.
The general solutions (\ref{eq:35}) and (\ref{eq:36}) have a very
simple form. It is important to observe that the Einstein-Maxwell
solutions (\ref{eq:35}) and (\ref{eq:36}) apply to both charged and
uncharged relativistic stars. We regain neutral exact solutions,
which may be possibly new, by setting $\alpha=0$.

\section{Known cases}
We may generate individual models for charged and uncharged stars
found previously from our general class of solutions. These can be
explicitly regained from the general series solution (\ref{eq:17})
or the elementary functions  (\ref{eq:35}) and (\ref{eq:36}). We
demonstrate that this is possible in the following cases:
\\\\
\textbf{Case I} : Thirukkanesh and Maharaj charged stars\\
If we set
$K=\frac{k-1}{k},~~\hat{\alpha}=\frac{\alpha}{k},~~c_{1}=d_{1}$
and $c_{2}=K^{-3/2}d_{2}$ then (\ref{eq:17}) can be written as
\bea \label{eq:37}
y&=&d_{1}\left[1+\sum_{i=0}^{\infty}\prod_{p=0}^{i}\frac{4p(p-1)-(K+\tilde{\alpha})}{2(p+1)(2p-1)}\left(\frac{1+x}{K}\right)^{i+1}\right]\n\\
&+&d_{2}\left(\frac{1+x}{K}\right)^{3/2}\left[1+\sum_{i=0}^{\infty}\prod_{p=0}^{i}\frac{(2p+3)(2p+1)-(K+\tilde{\alpha})}{(2p+5)(2p+2)}\left(\frac{1+x}{K}\right)^{i+1}\right]\eea
The Einstein-Maxwell  solution (\ref{eq:37}) corresponds to the
charged model of Thirukkanesh and Maharaj \cite{ThMa}. However
note that $k\neq 0$ and $k\neq 1$ in (\ref{eq:37}). In our wider
class of solutions (\ref{eq:17}) it is permitted that $k=0$; the
exact solution (\ref{eq:9a}) corresponds to the case $k=1$. The
Thirukkanesh-Maharaj charged stars contain neutron star models
found previously including the Durgapal and Bannerji model, and
they are consequently physically reasonable.\\\\
\textbf{Case II} : Hansraj and Maharaj charged stars\\
If we set $k=0$ then it is possible after some manipulation, to
write (\ref{eq:17}) in the form \bea \label{eq:41}
 y&=&c_{1}\left[1+\sum_{i=0}^{\infty}\frac{(-1)^{i}(2i+1)}{(2i+2)!}[(1-\alpha)(1
 +x)]^{i+1}\right]\n\\
&+&c_{2}(1+x)^{3/2}\left[1+\sum_{i=0}^{\infty}\frac{3(-1)^{i+1}(2i+4)}{(2i+5)!}[(1
-\alpha)(1+x)]^{i+1}\right]\n\\
&=&c_{1}\left[1-\sum_{i=0}^{\infty}(-1)^{i}\left(\frac{1}{(2i+2)!}-\frac{1}{(2i
+1)!}\right)[\sqrt{(1-\alpha)(1+x)}]^{2(i+1)}\right]\n\\
&+&\frac{3c_{2}}{(1-\alpha)^{3/2}}[\sqrt{(1-\alpha)(1+x)}]^{3}\n\\
&\times&\left[\frac{1}{3}-\sum_{i=0}^{\infty}(-1)^{i+1}\left(\frac{1}{(2i
+5)!}-\frac{1}{(2i+4)!}\right)[\sqrt{(1-\alpha)(1+x)}]^{2(i+1)}\right]\eea
It is interesting to observe that the last equation above can be
expressed in terms of trigonometric functions. Then it is easy to
show that (\ref{eq:41}) can be written in the form
\bea
\label{eq:41a}
 y&=&c_{1}[\cos\sqrt{(1-\alpha)(1+x)}+\sqrt{(1-\alpha)(1+x)}
 \sin\sqrt{(1-\alpha)(1+x)}]\n\\
&+&\frac{3c_{2}}{(1-\alpha)^{3/2}}[\sin\sqrt{(1-\alpha)(1+x)}-
\sqrt{(1-\alpha)(1+x)}\cos\sqrt{(1-\alpha)(1+x)}]\n\\
&=&[d_{1}+d_{2}\sqrt{(1-\alpha)(1+x)}]\sin\sqrt{(1-\alpha)(1+x)}\n\\
&+&[d_{2}-d_{1}\sqrt{(1-\alpha)(1+x)}]\cos\sqrt{(1-\alpha)(1+x)}\eea
where we have set $d_{1}=\frac{3c_{2}}{(1-\alpha)^{3/2}}$ and
$d_{2}=c_{1}$. The Einstein-Maxwell solution (\ref{eq:41a}) is the
same as the charged model of Hansraj and Maharaj \cite{HaMa}. Note
that our solution (\ref{eq:41a}) corrects a minor misprint in the
result of \cite{HaMa}. The Hansraj-Maharaj charged stars were
comprehensively studied in \cite{HaMa}; it was demonstrated that
their model produced a charged relativistic sphere that satisfied
all physical criteria. In particular the speed of sound was less
that the speed of light and causality is maintained.
\\\\
\textbf{Case III} : Finch and Skea neutron stars\\
If we set $k=0$ and $\alpha=0$, and follows the procedure outlined
for Case II, then  (\ref{eq:17}) becomes \bea \label{eq:42}
y&=&[d_{1}+d_{2}\sqrt{(1+x)}]\sin\sqrt{(1+x)}+[d_{2}-d_{1}\sqrt{(1+x)}]\cos\sqrt{(1+x)}\eea
where we have set $d_{1}=3c_{2}$ and $d_{2}=c_{1}$. Alternatively
we can obtain the result (\ref{eq:42}) directly from
(\ref{eq:41a}) with $\alpha=0$. The exact solution (\ref{eq:42})
is the neutron star model of Finch and Skea \cite{FiSk}. The
Finch-Skea model satisfies all the physical conditions for an
isolated spherically symmetric stellar source, and consequently
has been utilised by many researchers to model neutron
stars.\\\\
\textbf{Case IV} : Durgapal and Bannerji neutron stars\\
If we set $\alpha=0$ and $k=-\frac{1}{2}(n=0)$, then (\ref{eq:36})
becomes \be\label{eq:43}  y = c(1+x)^{3/2}+d(2-x)^{1/2}(5+2x)\ee
where we have set $c=A$ and $d=\frac{B}{3\sqrt{2}}$. The exact
solution (\ref{eq:43}) was first found by Durgapal and Bannerji
\cite{DuBa}. The Durgapal-Bannerji solution has been widely applied
as a relativistic model for neutral stars with superdense matter.

\section{Discussion}

We have found new solutions (\ref{20}) to the Einstein-Maxwell
system (\ref{eq:4}) by utilising the method of Frobenius for an
infinite series; a form for one of the gravitational potentials was
assumed and the electric field intensity was specified. These
solutions are given in terms of special functions. For particular
values of the parameters involved it is possible to write the
solution in terms of elementary functions: polynomials and products
of polynomials and algebraic functions. The electromagnetic field
may vanish in the general series solutions and we can regain
uncharged solutions. Thus our approach has the advantage of
necessarily containing a neutral stellar solution. The charged model
of Thirukkanesh and Maharaj \cite{ThMa}, the charged spheres of
Hansraj and Maharaj \cite{HaMa}, the uncharged neutron star model of
Finch and Skea \cite{FiSk}, and the uncharged superdense star of
Durgapal and Bannerji \cite{DuBa} are contained as special cases in
our class of solutions to the Einstein-Maxwell system. The simple
form of the solutions found facilitates the analysis of the physical
features of a charged sphere.

We are  now in a position to investigate graphically the behaviour
of the matter variables in the stellar interior
 for particular choices of the parameter values in (\ref{20}).
 We have generated figure 1 with the
 assistance of the software package Mathematica. For simplicity
 we make the choices $A=B=C=1$, $ k=\frac14$ and $\alpha = \frac32$,
 over the interval $0\leq r\leq 1$,
 to generate the relevant plots. In figure 1, plot A represents  the
 energy density $\rho$,  plot B represents the pressure $p$, and
 plot $C$ represents the electric field intensity $E^2$.
 It can easily be seen that these matter variables
 remain regular in the interior of the star for $0\leq r\leq 1$.
 We observe that the energy density and the pressure
 are positive and monotonically decreasing functions
in the interior of the star. The electric field intensity   is
positive and monotonically increasing for the interval of the plot.
Thus the quantities $\rho$, $p$ and $E^2$ are continuous, regular
and well behaved in the interior of the star. Consequently the
solutions found in this paper are likely to be useful in the
description of charged relativistic fluid spheres.

\begin{figure}[thb]
\vspace{1.5in} \includegraphics{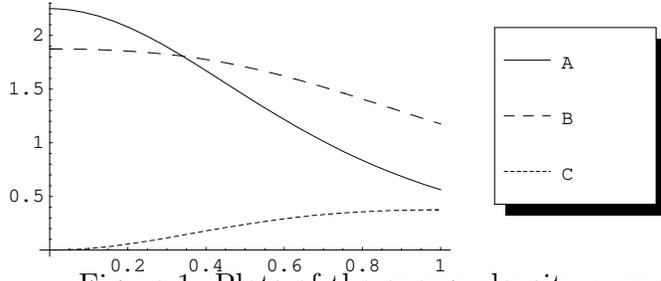} \caption{\label{plots1}
Plots of the energy density $\rho$, pressure $p$ and electric field
$E^2$.}
\end{figure}

\noindent{\large \bf Acknowledgements}\\

\noindent KK thanks the South Eastern University of Sri Lanka for
granting study leave and the National Research Foundation and the
University of KwaZulu-Natal for financial support. SDM
acknowledges that this work is based upon research supported by
the South African Research Chair Initiative of the Department of
Science and Technology and the National Research Foundation.
%

%
%
%
\end{document}